\documentclass[12pt]{article}%
\usepackage{amsmath}
\usepackage{graphicx}
\usepackage{amsfonts}
\usepackage{amssymb}%
\providecommand{\U}[1]{\protect\rule{.1in}{.1in}}
\begin{document}

\title{Friedel oscillations of Kondo impurities: A comparison.}
\author{Gerd Bergmann\\Department of Physics\\University of Southern California\\Los Angeles, California 90089-0484\\e-mail: bergmann@usc.edu}
\date{\today }
\maketitle

\begin{abstract}
Recently Affleck et al. derived the existence of Friedel oscillations in the
presence of a Kondo impurity. They supported their analytic derivation by
numerical calculations using Wilson's renormalization approach (NRG). In this
paper the size of the Friedel oscillations is calculated with the FAIR method
(Friedel Artificially Inserted Resonance) which has been recently developed.
The results of NRG and FAIR are compared. The development of Friedel
oscillations with a phase shift of $\pi/2$ outside of the Kondo radius is confirmed.

\end{abstract}

The properties of magnetic impurities in a metal is a fascinating problem
which was first studied by Friedel \cite{F28} and Anderson \cite{A31}. The
disappearance of the magnetic moment at \ low temperatures, the Kondo effect,
is one of the most intensively studied problems in solid state physics
\cite{K8}, \cite{D44}, \cite{H23}, \cite{M20}, \cite{A36}, \cite{G24},
\cite{C8}, \cite{H20}, \cite{W18}, \cite{N5}, \cite{N7}, \cite{B103},
\cite{W12}, \cite{A50}, \cite{S29}, \cite{N17}. In the last decade the Kondo
effect has experienced a renaissance. There is a growing interest in this
field \cite{G55}, extending from magnetic atoms on the surface of corrals
\cite{E18} to carbon nanotubes \cite{M93}, quantum dots \cite{G56},
\cite{A81}, \cite{A80}, \cite{A82}, \cite{B176}, \cite{S83}, \cite{G57} and
nanostructures \cite{P41}. There are still many open questions, particularly
the real-space form of the wave function and the resulting charge density and polarization.

Recently Affleck, Borda and Saleur \cite{A83} (ABS) investigated the formation
of Friedel oscillations in the vicinity of a Kondo impurity. Their result has
the form%
\begin{equation}
\rho_{Fr}\left(  r\right)  -\rho_{0}=\frac{C_{D}}{r^{D}}\left[  F\left(
\frac{r}{r_{K}}\right)  \cos\left(  2k_{F}r-D\frac{\pi}{2}\right)
-\cos\left(  2k_{F}r-D\frac{\pi}{2}\right)  \right] \label{fr_osc}%
\end{equation}
where $D$ is the dimension of the system, the coefficients $C_{D}$ have the
values $C_{1}=1/\left(  2\pi\right)  $, $C_{2}=1/\left(  2\pi^{2}\right)  $
and $C_{3}=1/\left(  4\pi^{2}\right)  ,$ and $r_{K}=\hbar v_{F}/k_{B}T_{K}$ is
the Kondo length. The function $F\left(  r/r_{K}\right)  $ is a universal
function which approaches the values $+1$ for $r/r_{K}<<1$ and $-1$ for
$r/r_{K}>>1$. (I skipped the phase shift $\delta_{P}$ due to potential scattering).

Besides the analytic derivation of (\ref{fr_osc}) ABS performed also numerical
calculations using Wilson's NRG approach \cite{W18}. In the NRG technique one
uses Wilson states with a logarithmic discretization of the conduction band
(which extends from $\left(  -1,+1\right)  $ (in units of the Fermi energy)
and has a constant density of states.)

ABS point out that Wilson energy states are not well suited for the
calculation of such fine effects as Friedel oscillations. The reason is that
the Wilson states average over a large number of original $\mathbf{k}$-states.
Therefore they use a modified method which was introduced earlier by Borda
(for details see \cite{B176}). With this approach they obtain a numerical
confirmation of the universal function $F\left(  r/\xi_{K}\right)  $.

The author has introduced during the past few years a new numerical approach
to the Kondo and the Friedel-Anderson impurity, the FAIR-method (Friedel
Artificially Inserted Resonance). It is based on the fact the $n$-electron
ground state of the Friedel Hamiltonian (consisting of an electron band and a
d-resonance) can be exactly expressed as the sum of two Slater states%
\begin{equation}
\Psi_{Fr}=Aa_{0}^{\dag}%
{\textstyle\prod\limits_{i=1}^{n-1}}
a_{i}^{\dag}\Phi_{0}+Bd^{\dag}%
{\textstyle\prod\limits_{i=1}^{n-1}}
a_{i}^{\dag}\Phi_{0}\label{psi_fr}%
\end{equation}
where $a_{0}^{\dag}$ is an artificial Friedel resonance state which determines
uniquely the full orthonormal basis $\left\{  a_{i}^{\dag}\right\}  $. An
extension of this ground state to the Friedel-Anderson and Kondo impurity
yields good numerical results. Recently this method was applied to calculate
the Kondo polarization cloud for those impurities \cite{B177}. Therefore the
author could not resist using the FAIR method to calculate the Friedel
oscillations of a Kondo impurity. As in the paper by ABS I am treating the
one-dimensional case. (The two- and three-dimensional treatments are
essentially identical). Also the calculation is restricted to zero impurity
scattering (the same as in ABS's calculations).

The construction of the Wilson states is essentially the same in NRG and FAIR.
One starts from an electron band with linear dispersion as shown in Fig.1. If
one follows Wilson and measures the energy in units of the Fermi energy and
the wave vector $\kappa=k/k_{F}$ in units of $k_{F}$ then the dispersion is
just
\[
\varepsilon_{\kappa}=\kappa-1
\]

This band contains a macroscopic number of states $\varphi_{\kappa}$ $\left(
\thickapprox10^{23}\right)  $. Wilson subdivided the positive and negative
half of the energy band by a factor $\lambda$. For example for $\lambda=2$ the
negative band is divided into energy cells $\mathfrak{C}_{0}=\left(
-1:-\frac{1}{2}\right)  ,$ $\mathfrak{C}_{1}=\left(  -\frac{1}{2}:-\frac{1}%
{4}\right)  ,$.. $\mathfrak{C}_{\nu}=\left(  \zeta_{\nu}:\zeta_{\nu+1}\right)
$, where $\zeta_{\nu}=-2^{-\nu}$. In each energy cell $\mathfrak{C}_{\nu}$ one
has $Z_{\nu}$ states $\varphi_{\kappa}$ with the energy $\left(
\kappa-1\right)  $. Assuming that each state $\varphi_{\kappa}\left(
r\right)  $ has the same interaction with the impurity then the states in the
cell $\mathfrak{C}_{\nu}$ are combined to a new state $\psi_{\nu}\left(
r\right)  $%
\[
\psi_{\nu}\left(  r\right)  =\frac{1}{\sqrt{Z_{\nu}}}%
{\textstyle\sum_{\kappa\in\mathfrak{C}_{\nu}}}
\varphi_{\kappa}\left(  r\right)
\]
This state represents the full interaction of all states $\varphi_{\kappa
}\left(  r\right)  $ in $\mathfrak{C}_{\nu}$ with the impurity. (Its
interaction is enhanced by $\sqrt{Z_{\nu}}$). The sub-Hilbert space of
$\mathfrak{C}_{\nu}$ contains $\left(  Z_{\nu}-1\right)  $ additional states
\[
\psi_{\nu}^{l}\left(  r\right)  =\frac{1}{\sqrt{Z_{\nu}}}%
{\textstyle\sum_{\kappa_{\mu}\in\mathfrak{C}_{\nu}}}
\varphi_{\kappa}\left(  r\right)  \exp\left(  i2\pi\frac{\mu l}{Z_{\nu}%
}\right)
\]
with $l$ and $\mu$ running from $0$ to$\ \left(  Z_{\nu}-1\right)  $ (Here
$\psi_{\nu}^{0}$ is identical to $\psi_{\nu}$). The states with $l>0$ have no
interaction with the impurity, but they have finite matrix elements
$\left\langle \psi_{\nu}^{l}\left\vert H_{0}\right\vert \psi_{\nu}^{l^{\prime
}}\right\rangle $ with the band energy Hamiltonian $H_{0}$. These matrix
elements are neglected in the NRG and the FAIR calculation. This is strictly
speaking only correct when all the original states $\varphi_{\kappa}\left(
r\right)  $ in a given energy cell $\mathfrak{C}_{\nu}$ have the same energy.
In Fig.1 the energy - wave vector diagram is shown. The dashed line gives the
linear dispersion relation. The darkly shaded square marks the energy
cell$\ \mathfrak{C}_{0}$. The thick zig-zag line gives the dispersion for
which the Wilson NRG would be exact.%

\begin{align*}
&
{\includegraphics[
height=3.3259in,
width=3.3574in
]%
{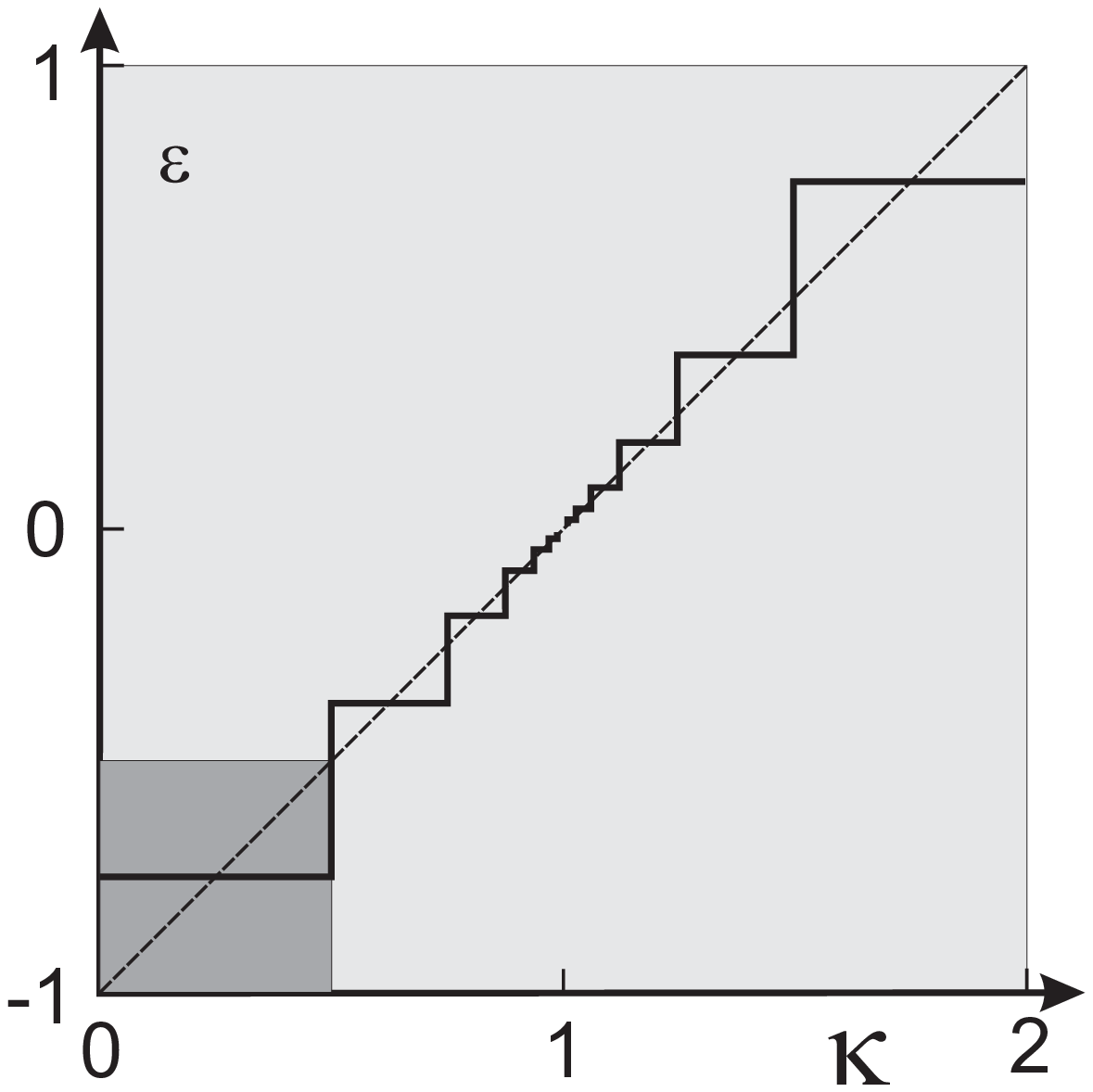}%
}%
\\
&
\begin{tabular}
[c]{l}%
Fig.1: The effective dispersion relation between energy $\varepsilon$ and\\
wave number $\kappa$ in Wilson's Hamiltonian.
\end{tabular}
\end{align*}

In the one-dimensional case the wave functions $\varphi_{\kappa}$ are cosine
and sine functions. (Only the cosine states interact with the impurity). Since
the energy $\zeta$ and the wave numbers $\kappa$ are given in normalized units
of $\varepsilon_{F}$ and $k_{F}$ it is natural to measure distances also in
reduced units. \textbf{Throughout this paper lengths are measured in units of
}$\mathbf{\lambda}_{F}\mathbf{/2}$\textbf{, for example the distance from the
impurity is denoted as }$\xi=r/\left(  \lambda_{F}/2\right)  $\textbf{.}

For the wave functions I use $\varphi_{\kappa}\left(  \xi\right)
=\sqrt{2/\Lambda}\cos\left(  \pi\kappa\xi\right)  $ in the range from
$-\Lambda/2$ to $+\Lambda/2$. (Here $\Lambda$ is given by the size $L$ of the
one-dimensional box, $\Lambda=L/\left(  \lambda_{F}/2\right)  .$ The number of
states $\varphi_{\kappa}\left(  \xi\right)  $ in the energy cell
$\mathfrak{C}_{\nu}$ is $\left(  \zeta_{\nu+1}-\zeta_{\nu}\right)  /\left(
2/\Lambda\right)  $

This yields for the wave function of the Wilson state $\psi_{\nu}\left(
\xi\right)  $
\begin{equation}
\psi_{\nu}\left(  \xi\right)  =\frac{2}{\sqrt{\left(  \zeta_{\nu+1}-\zeta
_{\nu}\right)  }}\frac{1}{\pi\xi}\sin\left(  \pi\xi\frac{\zeta_{\nu+1}%
-\zeta_{\nu}}{2}\right)  \cos\left(  \pi\xi\left(  1+\frac{\zeta_{\nu+1}%
+\zeta_{\nu}}{2}\right)  \right) \label{psi_nu}%
\end{equation}%
\[
\]

The exchange interaction is given by $H_{sd}$%

\begin{equation}
H_{sd}=v_{a}J%
{\textstyle\sum_{\alpha,\beta}}
\Psi_{\alpha}^{\dag}\left(  0\right)  \mathbf{\sigma}_{\alpha,\beta}%
\Psi_{\beta}\left(  0\right)  \cdot\mathbf{S}\label{Hsd}%
\end{equation}

where $v_{a}$ is the atomic volume and $J$ is the exchange interaction

The FAIR method yields for the Kondo ground state%
\begin{equation}
\Psi_{K}=\left[  A_{s,d}a_{0+\uparrow}^{\dag}d_{\downarrow}^{\dag}%
+A_{d,s}d_{\uparrow}^{\dag}a_{0-\downarrow}^{\dag}\right]  \overline
{\left\vert 0\right\rangle }+\left[  A_{d,s}a_{0-\uparrow}^{\dag}%
d_{\downarrow}^{\dag}+A_{s,d}d_{\uparrow}^{\dag}a_{0+\downarrow}^{\dag
}\right]  \overline{\overline{\left\vert 0\right\rangle }}\nonumber
\end{equation}%
\[%
\begin{array}
[c]{ccc}%
\overline{\left\vert 0\right\rangle }=%
{\textstyle\prod\limits_{i=1}^{n-1}}
a_{i+\uparrow}^{\dag}%
{\textstyle\prod\limits_{j=1}^{n-1}}
a_{j-\downarrow}^{\dag}\Phi_{0} &  & \overline{\overline{\left\vert
0\right\rangle }}=%
{\textstyle\prod\limits_{i=1}^{n-1}}
a_{i-\uparrow}^{\dag}%
{\textstyle\prod\limits_{j=1}^{n-1}}
a_{j+\downarrow}^{\dag}\Phi_{0}%
\end{array}
\]

The two FAIR states $a_{0+}^{\dag}$ and $a_{0-}^{\dag}$ are obtained by
optimizing the ground-state energy (see for example \cite{B153}, \cite{B177}).
They have the form
\[
a_{0\pm}^{\dag}=%
{\textstyle\sum_{\nu}}
\alpha_{0\pm}^{\nu}c_{\nu}^{\dag}%
\]
where $c_{\nu}^{\dag}$ are the creation operators of the Wilson states
$\psi_{\nu}$. (These two states are maximally coupled to the impurity. I
expect that they are closely related to Wilson's state $f_{0}$). The FAIR
states $a_{0\pm}^{\dag}$ determine uniquely the two bases $\left\{  a_{i\pm
}^{\dag}\right\}  $. The total charge density of the ground state $\Psi_{K}$
can be calculated from the ground state with the help of the individual wave
functions of the Wilson states.

To calculate the charge density of the Kondo ground state one has to calculate
the charge densities of all states $a_{i\pm}^{\dag}$ for $0\leq i<N/2$ and add
them according to the occupation of these states. This yields the main
contribution. But there are in addition two interference terms which yield
rather small contributions. The total charge density is then calculated at
average distances from the impurity of $2,4,8,..2^{l},...2^{20}$. This is done
in intervals of two wave lengths ($\Delta\xi=2$ or $\Delta r=\lambda_{F}$)
where a dense trace of the charge density as a function of $\xi$ is
calculated. For each trace the position of the charge minima and maxima is
determined as well as the amplitude.

In Fig.2 the numerical result for $F\left(  \xi/\xi_{K}\right)  $ is plotted.
It is the amplitude of the first term in equ. (\ref{fr_osc}), i.e. $C_{D}%
\cos\left(  2k_{F}r-D\frac{\pi}{2}\right)  /r^{D}$ (which in one dimensions is
equal to $\sin\left(  2\pi\xi\right)  /\left(  2\pi\xi\right)  )$. The
abscissa is the logarithm (base $2$) of the average distance $\xi$ from the
impurities. (The change in sign of $F\left(  \xi/\xi_{K}\right)
\ $corresponds to a jump in the phase by $\pi$).

The top curve with open circles is for $N=60$ Wilson states with a ratio
$\lambda=2$ between neighboring energies $\xi_{\nu}$. It confirms the
statement by ABS that these Wilson states are not well enough suited for the
calculation of small charge fluctuations on a length scale of the Fermi wave length.

To improve the calculation I tried a different path than ABS. In the FAIR
method the two essential states are the FAIR states $a_{0+}^{\dag}$and
$a_{0-}^{\dag}$. When they are known one can easily construct the full
orthonormal basis for each FAIR state (with the condition that $\left\langle
a_{i}^{\dag}\Phi_{0}\left\vert H_{0}\right\vert a_{j}^{\dag}\Phi
_{0}\right\rangle =0$ for $i,j>0$). This is a trivial calculation with
negligible computer time. Each $a_{0}^{\dag}$ is composed of Wilson states%
\[
a_{0}=%
{\textstyle\sum_{\nu=0}^{N-1}}
\alpha_{0}^{\nu}c_{\nu}^{\dag}%
\]
where $c_{\nu}^{\dag}$ is the creation operator of the Wilson state $\psi
_{\nu}$. The square $\left\vert \alpha_{0}^{\nu}\right\vert ^{2}$ gives the
occupation or contribution of the state $c_{\nu}^{\dag}$ or $\psi_{\nu}$ to
$a_{0}^{\dag}$. Since the state $\psi_{\nu}$ is constructed from the original
states $\varphi_{k}$ the ratio of $\left\vert \alpha_{0}^{\nu}\right\vert
^{2}$ divided by width of the energy cell, i.e. $p_{\nu}=\left\vert \alpha
_{0}^{\nu}\right\vert ^{2}/\left(  \zeta_{\nu+1}-\zeta_{\nu}\right)  $ gives
the relative contribution of the original states $\varphi_{\kappa}$ to the
FAIR state $a_{0}^{\dag}$. This state density $p_{\nu}$ corresponds to a
function $p\left(  \zeta\right)  $ with the condition $p\left(  \zeta\right)
=p_{\nu}$ for $\zeta_{\nu}<\zeta<\zeta_{\nu+1}$. The integral of $p\left(
\zeta\right)  $ over the energy is normalized%
\[
\int_{-1}^{+1}p\left(  \zeta\right)  d\zeta=%
{\textstyle\sum_{\nu=0}^{N-1}}
\int_{\zeta_{\nu}}^{\zeta_{\nu+1}}\frac{\left\vert \alpha_{0}^{\nu}\right\vert
^{2}}{\left(  \zeta_{\nu+1}-\zeta_{\nu}\right)  }d\zeta=%
{\textstyle\sum_{\nu=0}^{N-1}}
\left\vert \alpha_{0}^{\nu}\right\vert ^{2}=1
\]

For large $N$ the state density\ $p\left(  \zeta\right)  $ shows a smooth
dependence on the energy. (This is shown in Fig.4 for the example of a Friedel
resonance, see below). Therefore one can extrapolate $p\left(  \zeta\right)  $
from a finite representation $N$ and construct the FAIR state for twice the
number of states. The most natural way to double the number of Wilson states
is to set a new energy ratio $\zeta_{\nu}^{\prime}/\zeta_{\nu+1}^{\prime}=$
$\lambda^{\prime}=\sqrt{\lambda}$. In this way the FAIR states and their full
bases are obtained for $N=120$ (with a new $\lambda=\sqrt{2}$). It turns out
that the new solution for $\Psi_{K}$ is better than the old one (for $N=60)$
because its ground-state energy is lowered. The second curve in Fig.2 (open
triangles) shows the numerical result for the function $F\left(  \xi/\xi
_{K}\right)  $. It clearly improves the result although it does not yet reach
the value $-1$ for large $\xi$. A second doubling of $N$ yields $N=240$ and
$\lambda=\sqrt[4]{2}$. The ratio: width of an energy cell divided by the
average energy is given by $2\left(  \lambda-1\right)  /\left(  \lambda
+1\right)  .$ For the usual Wilson states with $\lambda=2$ this ratio is
$2/3$. After two additional splitting of the states this ratio becomes about
$\allowbreak0.173$. Now the energy states are much closer together. For
$N=240$ the numerical result for $F\left(  \xi/\xi_{K}\right)  $ is shown in
the bottom curve of Fig.2. It finally shows the two limiting values of $+1$ at
short distances and $-1$ at large distances in agreement with the theoretical prediction.%

\begin{align*}
&
{\includegraphics[
height=2.5571in,
width=3.1133in
]%
{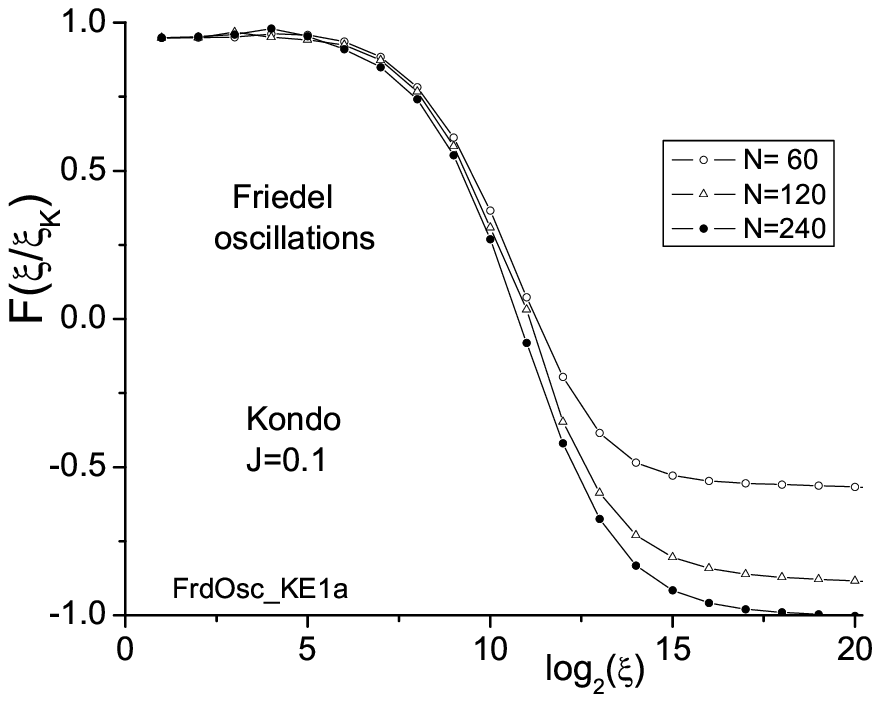}%
}%
\\
&
\begin{tabular}
[c]{l}%
Fig.2: The numerical result for the function $F\left(  \xi/\xi_{K}\right)  $
in equ. (\ref{fr_osc})\\
for different ratios of neighboring energies $\zeta_{\nu}$. For the number of
Wilson\\
states $N=60$ the ratio is $\lambda=2$, for $N=120$ it is $\lambda=\sqrt{2} $
and for $N=240$\\
it is $\lambda=\sqrt[4]{2}$.
\end{tabular}
\end{align*}

Since the function $F\left(  \xi/\xi_{K}\right)  $ is a universal function a
comparison of Kondo impurities with different Kondo energies is appropriate.
ABS performed a large number of calculations. I want to show that the FAIR
method yields the same universality. In the past we investigated Kondo
impurities with different exchange interactions and therefore different Kondo
energies \cite{B153}. Here I consider the two examples with $J=0.1$ and
$J=0.08.$ One obtains the Kondo lengths from the corresponding the Kondo
energies $\varepsilon_{K}=2.35\times10^{-5}$ and $1.37\times10^{-6}$.

In Fig.2 we plotted the amplitude of the first term in equ.(\ref{fr_osc}) to
extract the function $F\left(  \xi/\xi_{K}\right)  $. Now we determine the
actual amplitude $A\left(  \xi/\xi_{K}\right)  $ of the Friedel oscillation
which is given by
\[
\rho\left(  \xi\right)  -\rho_{0}=\left[  F\left(  \xi/\xi_{K}\right)
-1\right]  \frac{\cos\left(  2\pi\xi-\frac{\pi}{2}\right)  }{2\pi\xi
}=-A\left(  \xi/\xi_{K}\right)  \frac{\cos\left(  2\pi\xi-\delta\right)
}{2\pi\xi}%
\]

In Fig.3 this amplitude $A\left(  \xi/\xi_{K}\right)  =\left[  1-F\left(
\xi/\xi_{K}\right)  \right]  $ is plotted for the two different Kondo energies
as a function of $\log_{2}\left(  \xi/\xi_{K}\right)  $. Since the abscissa
has a logarithmic scale the curves are just shifted by $\log_{2}\left(
\xi_{K}\right)  .$ Although the Kondo energies differ by roughly a factor of
five the two curves are essentially identical. Therefore the amplitude\ of the
Friedel oscillation is universal (for the symmetric case and a linear
dispersion relation). The numerical calculation yields a phase shift of
$\pi/2$ for distances larger than 2 Fermi wave lengths. For most of the region
($6\leq\xi/\xi_{K}\leq16)$ the relative deviation is less than $10^{-3}$.

ABS give two asymptotic function for $\xi<<\xi_{K}$ and $\xi>>\xi_{K}$. These
functions which are shown as dotted curves in Fig.3, have been shifted by
$0.5$ to the left (see upper scale). This means that I use a slightly
different definition of the Kondo energy than ABS, which yields a different
Kondo length. The ratio between our Kondo energies is $2^{0.5}\thickapprox
1.4$. This is not surprising because ABS use Wilson's definition which
requires the NRG calculation whereas we define the Kondo energy as the energy
difference between triplet and singlet state \cite{B153}. ABS obtained from
their numerical calculation that the function $F\left(  \xi/\xi_{K}\right)  $
crosses zero at $\left(  \xi/\xi_{K}\right)  \thickapprox0.12\pm0.02$. In
Fig.3 this point lies at $2^{-2.9}=0.13$.%

\begin{align*}
&
{\includegraphics[
height=3.3723in,
width=4.0473in
]%
{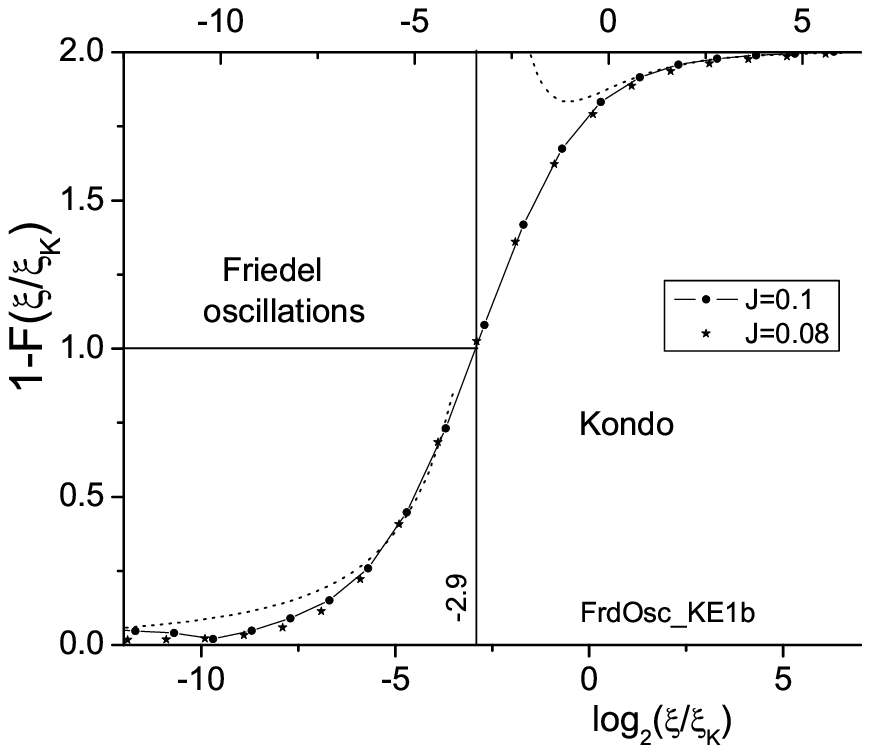}%
}%
\\
&
\begin{tabular}
[c]{l}%
Fig.3: The amplitude of the Friedel oscillation $\left[  1+F\left(  \xi
/\xi_{K}\right)  \right]  $ is\\
plotted versus $\xi/\xi_{K}$ for two different Kondo energies.
\end{tabular}
\end{align*}

It is worthwhile mentioning that in this calculation the Kondo ground state is
calculated once. From this ground state one obtains the Friedel oscillation
for all distances $\xi$. ABS had to perform a new NRG calculation for each
individual distance $\xi$ (which actually represent states with slightly
different ground-state energies).

One essential step in calculating the Friedel charge oscillations was the
repeated doubling of the number of Wilson states. This was possible because
the compositions of the FAIR states $a_{0+}^{\dag}$ and $a_{0-}^{\dag}$ are
essentially given by the state density functions $p_{+}\left(  \zeta\right)  $
and $p_{-}\left(  \zeta\right)  $. This shall be demonstrated for the FAIR
state of a Friedel (resonance) impurity.

We take the Wilson electron band ranging from $\left(  -1:+1\right)  $ with
constant density of states (equal to $1/2$) and a d resonance at the Fermi
energy ($E_{d}=0$) with a s-d matrix element $\left\vert V_{sd}\right\vert
^{2}=0.1$. The ground state is given in (\ref{psi_fr}) with the FAIR state
$a_{0}^{\dag}$.

In Fig.4 the values $p_{\nu}=\left\vert a_{0}^{\nu}\right\vert ^{2}/\left(
\zeta_{\nu+1}-\zeta_{\nu}\right)  $ are plotted at the discrete energies
$\left(  \zeta_{\nu+1}+\zeta_{\nu}\right)  /2.$ The full circles are the
values for $N=24$, $\lambda=2$. Only a small fraction of the whole energy band
is included in the figure to show an optimal section. The full triangles show
the values for $N=48$, $\lambda=\sqrt{2}$ while the stars represent the values
of $p_{\nu}$ for $N=96$, $\lambda=\sqrt[4]{2}$. For each $N$ the optimal
ground state has been derived. One recognizes that the points follow indeed a
common function $p\left(  \zeta\right)  $. This justifies the doubling of the
Wilson states since the new $p_{\nu}$ values can be obtained by inter- and
extrapolation. This yields the composition of the new FAIR state after the
doubling of the number of Wilson states $N$.
\begin{align*}
&
{\includegraphics[
height=2.709in,
width=3.1681in
]%
{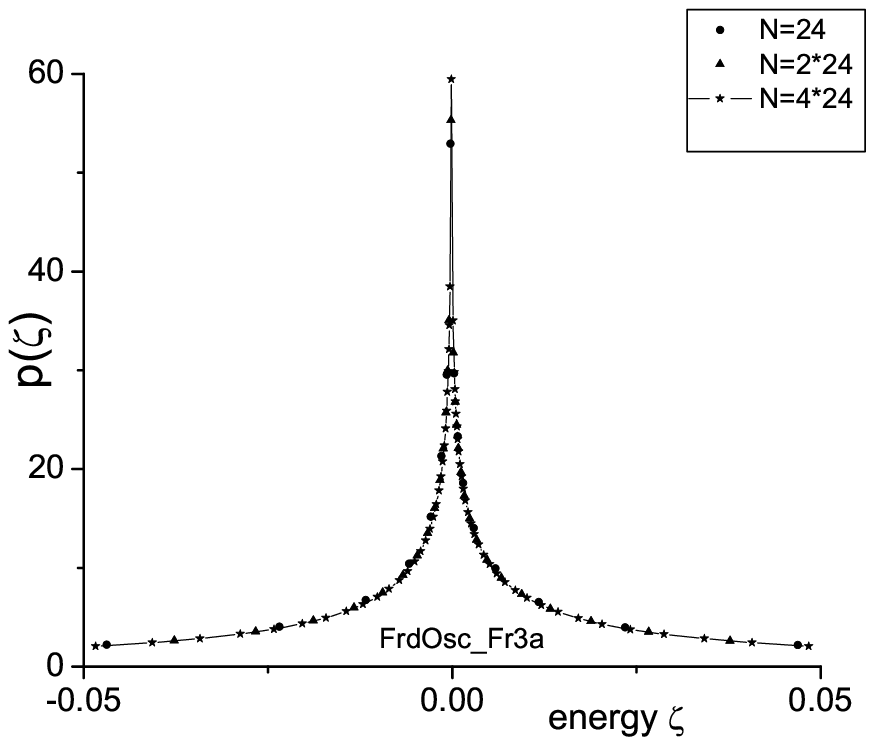}%
}%
\\
&
\begin{tabular}
[c]{l}%
Fig.4: The function $p\left(  \zeta\right)  \thickapprox\left\vert \alpha
_{0}^{\nu}\right\vert ^{2}/\left(  \zeta_{\nu+1}-\zeta_{\nu}\right)  $ for
$N=24,48,96$\\
with $\lambda=2,\sqrt{2}$ and $\sqrt[4]{2}$.
\end{tabular}
\end{align*}

As an after thought of this consideration I calculated the Friedel oscillation
for Friedel impurities with the d-resonance energy at the Fermi level,
$E_{d}=0$, and very small resonance width $\Delta=\pi\left\vert V_{sd}%
\right\vert ^{2}\rho$ where $\rho=1/2$ represents the density of states for
the Wilson spectrum. By using rather small values for $\left\vert
V_{sd}\right\vert ^{2}$= $0.5\times10^{-4}$, $1.0\times10^{-4}$ and
$2.0\times10^{-4}$ one can evaluate the amplitude of the Friedel oscillations
as a function of $\xi$.

The numerical calculation yields a charge oscillation $\Delta\rho_{Fr}%
=A_{Fr}\left(  \xi\right)  \cos\left(  2\pi\xi-\delta_{Fr}\right)  /\left(
2\pi\xi\right)  $. The phase shift $\delta_{Fr}$ in the range $3\leq\log
_{2}\xi\leq19$ is equal to $\pi/2$ within $1\%$ accuracy. For shorter
distances shorter than the Friedel coherence length ($\xi_{Fr}\thickapprox
2/\left\vert \pi V_{sd}\right\vert ^{2}$) the amplitude is strongly
suppressed. The amplitude is also universal since the three curves in Fig.5
would perfectly coincide if plotted as a function of $\left(  \xi/\xi
_{Fr}\right)  $. This behavior is quite analogous to the Friedel oscillations
of a Kondo impurity. However, the functional dependence and the long-distance
amplitude differ.%

\begin{align*}
&
{\includegraphics[
height=3.2312in,
width=3.907in
]%
{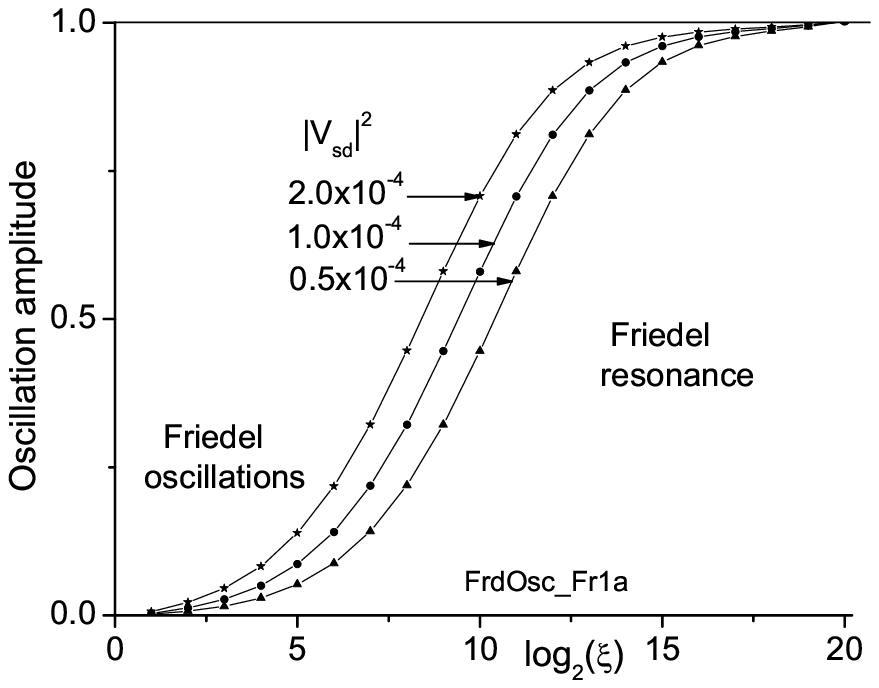}%
}%
\\
&
\begin{tabular}
[c]{l}%
Fig.5: The amplitude of the Friedel oscillation for Friedel\\
impurities with s-d hopping matrix elements of\\
$\left\vert V_{sd}\right\vert ^{2}=2\times10^{-4},1\times10^{-4}$ and
$.5\times10^{-4}.$ The Friedel\\
oscillations develop only for distances larger than the\\
Friedel length $\xi_{Fr}$.
\end{tabular}
\end{align*}

To conclude, the goal of this paper has been to reproduce ABS result for the
Friedel oscillations of a Kondo impurity with the FAIR method. This is a
rather delicate problem because electron densities have to be calculated on a
spatial scale much smaller than the Fermi wave length $\lambda_{F}$ over
distances of up to $10^{6}\lambda_{F}$ with a relative accuracy of
$\lambda_{F}/r$, i.e. $10^{-6}$ or better. The numerical results of the FAIR
method have about the same quality as the NRG method.

\end{document}